\documentclass[]{interact}

\usepackage{epstopdf}
\usepackage[caption=false]{subfig}
\usepackage{hyperref}
\usepackage[linesnumbered,ruled,vlined]{algorithm2e}
\usepackage[longnamesfirst,sort]{natbib}
\bibpunct[, ]{(}{)}{;}{a}{,}{,}
\renewcommand\bibfont{\fontsize{10}{12}\selectfont}

\theoremstyle{plain}

\theoremstyle{definition}

\theoremstyle{remark}

\begin{document}

\title{A Dynamic Topic Identification and Labeling Approach of COVID-19 Tweets}

\author{
\name{Khandaker Tayef Shahriar\textsuperscript{a}\thanks{*Correspondence: iqbal@cuet.ac.bd (Iqbal H. Sarker)}, Iqbal H. Sarker\textsuperscript{a,*}, Muhammad Nazrul Islam\textsuperscript{b}, Mohammad Ali Moni\textsuperscript{c,d}}
\affil{\textsuperscript{a}Department of Computer Science \& Engineering, Chittagong University of Engineering \& Technology, Chittagong-4349, Bangladesh. \\
\textsuperscript{b}Department of Computer Science and Engineering, Military Institute of Science and Technology, Dhaka-1216, Bangladesh. \\
\textsuperscript{c}WHO Collaborating Centre on eHealth, UNSW Digital Health, Faculty of Medicine, University of New South Wales, Sydney, NSW 2052, Australia. \\
\textsuperscript{d}Healthy Ageing Theme, Garvan Institute of Medical Research, Darlinghurst, NSW 2010, Australia,}
}

\maketitle

\begin{abstract}
This paper formulates the problem of \textit{dynamically identifying key topics with proper labels} from COVID-19 Tweets to provide an overview of wider public opinion. Nowadays, social media is one of the best ways to connect people through Internet technology, which is also considered an essential part of our daily lives. In late December 2019, an outbreak of the novel coronavirus, COVID-19 was reported, and the World Health Organization declared an emergency due to its rapid spread all over the world. The COVID-19 epidemic has affected the use of social media by many people across the globe. Twitter is one of the most influential social media services, which has seen a dramatic increase in its use from the epidemic. Thus \textit{dynamic extraction of specific topics with labels} from tweets of COVID-19 is a challenging issue for highlighting conversation instead of manual topic labeling approach. In this paper, we propose a \textit{framework} that automatically identifies the key topics with labels from the tweets using the top Unigram feature of aspect terms cluster from Latent Dirichlet Allocation (LDA) generated topics. Our experiment result shows that this \textit{dynamic topic identification and labeling} approach is effective having the accuracy of 85.48\% with respect to the manual static approach.
\end{abstract}


\section{Introduction}

Social media platforms play an important role in extreme difficulties. People use the Internet connection channels to share ideas and provide feedback to others related to coping with disaster response and generate information and opinions. As a social media platform, Twitter greatly impacts daily lives regardless of geographical location. At the end of December 2019, COVID-19 was reported that was produced by a novel coronavirus outbreak~\citep{Malta}. The World Health Organization announced a state of emergency, due to the rapid spread of the coronavirus. COVID-19 pandemic has a great impact on the uses of Twitter. So it is very important to know the topic of tweets that users are generating on this platform regularly to get ideas about the crisis, people’s needs, and steps to recover the effect of the pandemic situation. However, it is a challenging issue to extract meaningful topics automatically instead of manual topic labeling approach~\citep{Sarker2}. COVID-19-related tweets from Twitter can be considered helpful in providing meaningful topics to better understand the ideas and highlight the user conversations. Therefore, in the proposed work, we focus on the dynamic extraction of key topics associated with proper topic labels from people's tweets at the COVID-19 epidemic event.

Let’s consider a Twitter dataset containing information about COVID-19 related issues. It is very cumbersome to analyze all the tweets and unveiling the internal gist manually. Hence, identifying internal topics automatically of all tweets will help reformists in relevant departments to implement obligate steps to minimize the detrimental outcome of the pandemic situation. Our goal is to propose an effective approach to automatically identify the topic associated with the proper label of tweets to depict the valid issues of the pandemic.

Latent topic models are effective ways to extract latent semantic data from the corpus~\citep{Blei,Deer}. Among these types, Latent Dirichlet Allocation (LDA)~\citep{Blei} seems to be the most effective for topic modeling. The document is considered as a combination of hidden topics by LDA modeling. As a result, the document is allowed to be on multiple topics. Labeling those topics with an appropriate attribute tag is a great challenge. However, Unigram is a type of probabilistic language model that exhibits the context with a well-implied space-time tradeoff. Moreover, Aspect terms depict sentiment features of an entity from user-generated texts~\citep{WangCoupled}. In the proposed framework, we use the top Unigram feature of each aspect terms cluster of LDA generated topics to label those topics for understanding ideas and highlighting user conversation. The summary of the main contributions of this paper is given below:
\begin{itemize}
\item We effectively use aspect terms of tweets to generate meaningful topic labels.
\item We propose a framework that is capable of extracting topics from COVID-19 Tweets and dynamically label them rather than following manual approach.
\item We have conducted a range of experiments to show the effectiveness of our topic identification and labeling approach.
\end{itemize}

The rest of this paper is organized as follows - Section 2 reviews works related to topic identification. The methodology of the proposed model is presented in section 3. After that, the experimental results are discussed in section 4. Finally, we conclude our work and discuss the direction of future research.

\section{Related Work}

COVID-19-related tweets can be considered helpful in delivering meaningful topics to better understand ideas and highlight human conversations. The LDA model for topic extraction has been used by many researchers. Patil et al.~\citep{Patil} presented a topic extraction model of people reviews using the frequency-based technique without improvising no robust technique for topic labeling. Asmussen et al.~\citep{Asmussen} proposed a paper by using the topic modeling approach for researchers but the topic labeling depends on the researcher without any automatic method. Wang et al.~\citep{Wang} presented a topic modeling system that is efficient in alleviating the data sparsity problem without identifying key topics with a specific label of tweets. Ramos et al.~\citep{Ramos} extracted topics by utilizing Negative Matrix Factorization (NNMF) and Term Frequency Inverse Document Frequency (TFIDF) but a popular model like Latent Dirichlet Allocation (LDA) could be implemented for topic extraction. Zhu et al.~\citep{Zhu} presented that the number of texts on topics changes with time,  but topic labeling was done manually. Lu et al.~\citep{Lu} explored the topic modeling impact for information retrieval but there is a function in which only coarse matching is required and training data is sparse. Buntime et al.~\citep{Buntine} developed a data acquisition system based on the technique of
a hierarchical topic modeling for retrieving documents that are most closely related to
the query given. Md. Shahriare Satu et. al.~\citep{Satu} proposed TClustVID that extracts meaningful sentiments but some explanation of topics manually may misinterpret in the topic modeling. Kee et al.~\citep{Kee} explained higher-order arbitrary topics extracted by the LDA model but topic evaluation to clear Clear Collective Themes was only 61.3\%.

The conclusion from the above works shows that most of the works followed the manual topic labeling approach, which might be a significant issue considering diverse human interpretations in the real world. Hence, a new and effective approach can be considered for automatic topic labeling. Our further work on this paper is based on developing a dynamic topic identification and labeling framework of COVID-19 related tweets.
\section{Methodology}
In this section, we present our framework as shown in Figure 1. Text preprocessing is the most essential step for analyzing tweets. The overall process for dynamic topic identification and labeling from preprocessed text is set out in Algorithm 1. After preprocessing of Twitter dataset our approach consists of several processing steps discussed below.
\subsection{Aspect Terms Extraction} 
We only consider nouns and noun phrases~\citep{Liu} as aspect terms of a sentence. Aspect terms are often regarded as objects of verbs or associated with modifiers that express sentiment. Aspect terms generally refer to portions of a sentence that notice an aspect of the product, event, entity, etc. Parts of speech tagging is an effective approach for separating noun chunks from sentences. Table 1. exhibits the examples of aspect terms from sample Tweets.
\begin{figure}
\includegraphics[width=\textwidth,height=5cm]{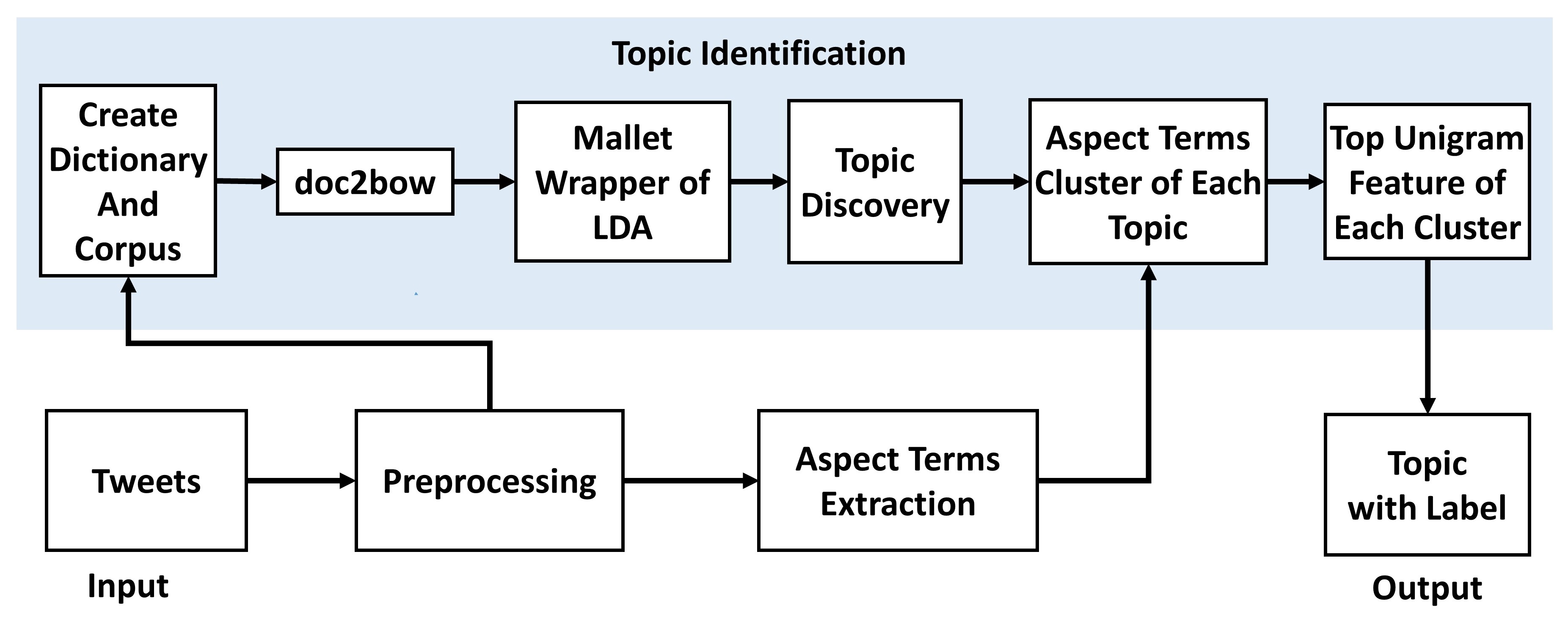}
\caption{Proposed framework for dynamic topic identification and labeling}
\end{figure}
\begin{algorithm}[H]
\DontPrintSemicolon
  
  \KwIn{T: number of Preprocessed Tweets in dataset}
  \KwOut{Topic Label ($T_{Label}$).}

  \For{\textbf{each} t $\in$ \{1,2,...,T\} }{
  \tcp{Corpus Development}
  $ C \leftarrow Create\_Corpus(t);$
  
    \tcp{Aspect Terms Extraction}
  $A_T \leftarrow Aspect\_Terms(t);$
 }
  \tcp{Topic Discovery}
  $K\thicksim Mallet(LDA(Doc2bow(C)));$
  
  \For{\textbf{each} k $\in$ \{1,2,...,K\} }{
  
  \For{\textbf{each} t $\in$ \{1,2,...,T\} }{
  $k_{dominant,t} \thicksim dominant\_topic(t,k);$ 
  
  \tcp{Aspect Terms Cluster from Topic} $C_A \thicksim Cluster( A_T\rightarrow k_{dominant,t});$}}
  \For{\textbf{each} k $\in$ \{1,2,...,K\} }{$T_{Label}\thicksim max\_count(Top\_Unigram(C_A\rightarrow k));$}
  
\caption{Topic Identification and Labeling}
\end{algorithm}
\begin{table}[]
\caption{An example of aspect terms}
\begin{center}
\begin{tabular}{|c|c|}
\hline
\textbf{Sample Tweet}                               & \textbf{Aspect Terms} \\ \hline
\begin{tabular}[c]{@{}c@{}}Hi coronavirus. Thanks for making me do more \\ online shopping.\end{tabular} &
  \begin{tabular}[c]{@{}c@{}}coronavirus, thanks, \\ shopping\end{tabular} \\ \hline
Great place to relax and enjoy your dinner     & place, dinner         \\ \hline
This staff should be fired.                    & staff                 \\ \hline
Was at the supermarket today. Didn't buy soap. & supermarket, soap     \\ \hline
\end{tabular}
\end{center}
\end{table}
\subsection{Topic Identification} Since LDA is a popular topic modeling technique~\citep{Sarkeretal}, the challenge is how to produce the highest quality of clear, categorized, and meaningful topic labels. This depends largely on the quality of the text preprocessing and the strategy for obtaining the total number of topics. The topic identification process contains several steps as discussed below:
\begin{enumerate}
\item[{\it 1)}]{\it Creating dictionary and corpus:}

A corpus can be referred to as an arbitrary sample of language and a dictionary helps to create a systematic account of the lexicon of a language. We create the dictionary of words and corpus from the preprocessed text. 
\item[{\it 2)}]{\it Creating a BoW corpus:}

In every document, the word id and its frequency is contained by the corpus. Doc2bow converts documents into Bag of Words (BoW) format. Assuming each word as a tokenized and normalized string.
\item[{\it 3)}]{\it Topic discovery:}

We transfer the BoW corpus to the Mallet wrapper of LDA~\citep{Blei}. LDA is a well-known topic modeling algorithm that allows viewing sets, defined by invisible groups to explain why certain parts of the data are similar. It illustrates documents as a combination of topics that spit out words with certain probabilities. Mallet has an active LDA startup. It offers a better division of topics and is known to run faster. We discover a model with 20 topics itself as shown in Figure 2. In Figure 2 we visualize the topics in the two-dimensional plane as circles whose centers are defined by the calculation of the Jensen–Shannon divergence~\citep{Fuglede} between topics. Then we project the inter-topic distances by using multidimensional scaling onto two dimensions. We use the areas of circles to encode the overall prevalence of each topic. We then find the dominant topic of each Tweet that is the topic to which a particular Tweet belongs.
\item[{\it 4)}]{\it Generation of Aspect Terms Cluster:}

Aspect terms of tweets corresponding to each LDA-generated topic are clustered. In this way, we get 20 aspect terms cluster. The aspect term carries the sentiment of a document. 
\item[{\it 5)}]{\it Labeling topic with top Unigram feature:}

A Unigram is a special form of n-gram, where n is 1. They are often used in natural language processing, mathematical text analysis, and cryptography for the control and use of ciphers and codes. The top unigram feature of each aspect terms cluster is considered to be the topic label of that particular topic. If two topics contain the same Top Unigram feature, then the topic which has the highest count is considered as topic label and in case of another topic, next Top Unigram feature is used for labeling.  An aspect term tag is used to identify a topic because that feature word reflects an attribute, which could be the topic we are talking about in a particular tweet.
\end{enumerate}
Within the section of the Methodology of this work, we try to identify the key topics generated from the Tweets and assign an appropriate topic label as shown in Table 2. To determine the topic of a Tweet we find the topic number that has a greater percentage on that Tweet.
\begin{figure}
\begin{center}
\includegraphics[width=25.5pc,height=8cm]{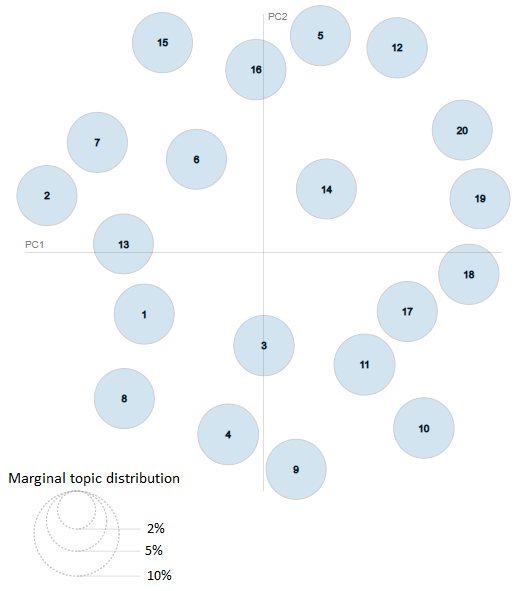}
\caption{Intertopic Distance Map (via multidimensional scaling)}
\end{center}
\end{figure}
\section{Experiments}
\begin{figure}
\begin{center}
\includegraphics[width=25.5pc,height=6cm]{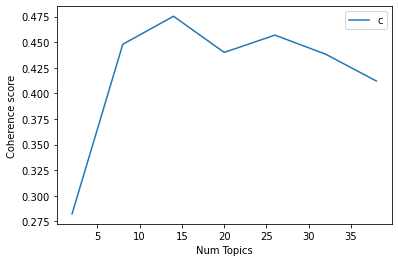}
\caption{Selecting the optimal number of LDA topics}
\end{center}
\end{figure}
\subsection{Dataset}
Twitter dataset was collected from the website at \url{https://www.kaggle.com/datatattle/covid-19-nlp-text-classification}. Dataset contains two csv files namely Corona\_NLP\_train.csv and Corona\_NLP\_test.csv. Corona\_NLP\_train.csv file contains 41,157 and Corona\_NLP\_test.csv contains 3,798 Covid-19 related tweets.
\subsection{Data Preprocessing}
Due to the freestyle of writing of posts on Twitter, a huge amount of noisy data is contained by the corpus. Hence, the elimination of noisy and ill-formatted data from the corpus is an essential task for us. Our preprocessing step contains the following functions:
\begin{enumerate}
\item[{\it 1)}]{\it Transforming words into lowercase:}
All words in the corpus are transformed in lowercase.

\item[{\it 2)}]{\it Replacing links with empty string:}
All the URL’s and hyperlinks are replaced with empty string in the tweets.

\item[{\it 3)}]{\it Replacing mentions with empty string:}
Usually, people post tweets by mentioning a user with the @. In our work we remove the mentions.

\item[{\it 4)}]{\it Replacing hashtags with empty string:}
Hashtags are identified and eliminated.

\item[{\it 5)}]{\it Dealing with contractions:}
Generally users on social media use common contractions in English. We handle this by utilizing regular expression.

\item[{\it 6)}]{\it Replacing punctuation with space:}
Punctuation available in tweets has no importance. We withdraw punctuation symbols such as ;, \&, -, \_ etc. in this step.

\item[{\it 7)}]{\it Striping space from words:}
Leading and ending spaces are removed.

\item[{\it 8)}]{\it Removing words less than two characters:}
All words whose length is less than two are removed.

\item[{\it 9)}]{\it Removing stop words:}
Stop words like articles, pronouns, prepositions etc. do not provide any information. These stop words are removed.

\item[{\it 10)}]{\it Non-English words and Unicode handling:}
Tweets should be in standard form. So, to clean the dataset at the primary step, the tweets that are not in English and Unicode like ‘‘\textbackslash u018e’’ be removed which are caused by miscellanies in the crawling process.
\end{enumerate}
\subsection{Optimal number of LDA topic selection:}
Our way of finding the total number of topics is to create multiple LDA models with different values of the number of topics (k) and then choose the one that offers the highest coherence value (CV). Selecting the 'k' that marks the end of the rapid growth of topic coherence usually provides meaningful and interpretive topics as shown in Figure 3.
\begin{figure}
\begin{center}
\includegraphics[width=\textwidth]{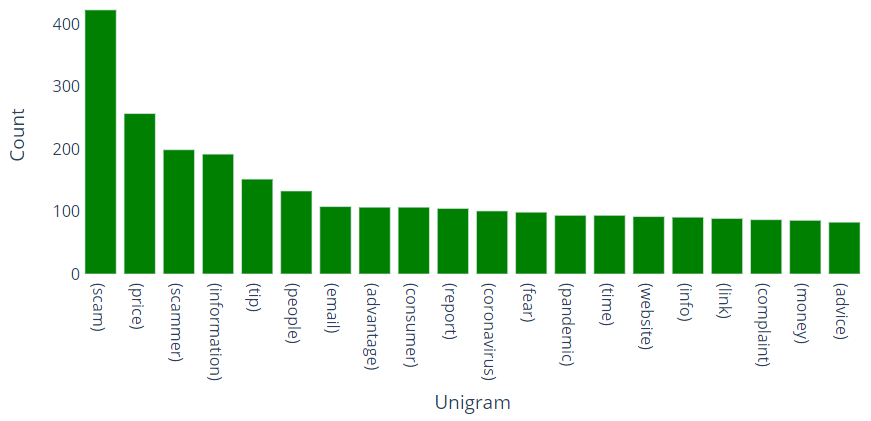}
\caption{Top 20 unigrams of topic no. 7 from COVID-19 tweets}
\end{center}
\end{figure}
\begin{figure}
\begin{center}
\includegraphics[width=\textwidth]{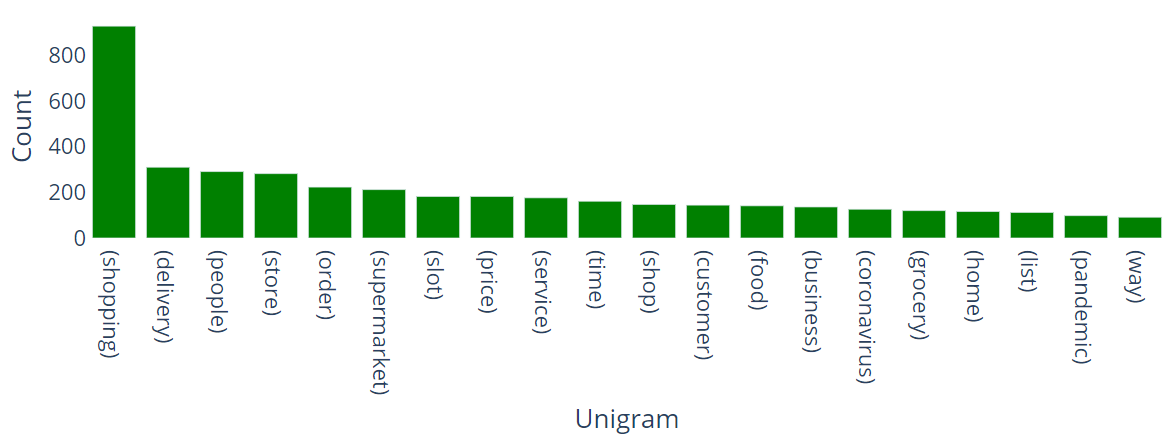}
\caption{Top 20 unigrams of topic no. 6 from COVID-19 tweets}
\end{center}
\end{figure}
\begin{table}[]
\caption{Example of Topics Detected on Tweets}
\begin{center}
\begin{tabular}{|c|c|c|}
\hline
\textbf{Sample Tweet} &
  \textbf{\begin{tabular}[c]{@{}c@{}}Topic\\ No.\end{tabular}} &
  \textbf{\begin{tabular}[c]{@{}c@{}}Detected\\ Topic Label\end{tabular}} \\ \hline
\begin{tabular}[c]{@{}c@{}}Lines at the grocery store have been unsure,\\ but is eating out  a safe alternative?\end{tabular} &
  2 &
  Store \\ \hline
\begin{tabular}[c]{@{}c@{}}For those who aren't struggling, please\\ consider donating to a food bank or a\\ nonprofit. The demand for these services\\ will increase as COVID-19 impacts jobs,\\ and people's way of life.\end{tabular} &
  12 &
  Demand \\ \hline
\begin{tabular}[c]{@{}c@{}}We're here to provide a safe shopping\\ experience for our customers and a\\ healthy environment for our associates\\ and community!\end{tabular} &
  6 &
  Shopping \\ \hline
\begin{tabular}[c]{@{}c@{}}My thoughts on impacts of coronavirus\\ on food markets\end{tabular} &
  4 &
  Consumer \\ \hline
\begin{tabular}[c]{@{}c@{}}I love this ?? all grocery stores should have\\ a set hour (preferably in the morning) for the\\ elderly population or those with compromised\\ immune systems to use the store\end{tabular} &
  8 &
  Time \\ \hline
Cleanshelf supermarket Sanitizers 19 &
  17 &
  Sanitizer \\ \hline
\begin{tabular}[c]{@{}c@{}}If you find yourself needing work, food\\ delivery driver seems like it is going to be\\ in high demand\end{tabular} &
  19 &
  Worker \\ \hline
\begin{tabular}[c]{@{}c@{}}While coronavirus (COVID-19) has sparked\\ some consumer concern over fresh foods,\end{tabular} &
  18 &
  Food \\ \hline
\begin{tabular}[c]{@{}c@{}}i'm so affraid of covid-19 that i'll spend my\\ whole evening in a crowded supermarket\end{tabular} &
  3 &
  Supermarket \\ \hline
\begin{tabular}[c]{@{}c@{}}Huge increases in online shopping have been\\ reported. What are your go-to apps during the\\ pandemic?\end{tabular} &
  6 &
  Shopping \\ \hline
\begin{tabular}[c]{@{}c@{}}Working in a grocery store right now\\ sounds like hell.\end{tabular} &
  2 &
  Store \\ \hline
\begin{tabular}[c]{@{}c@{}}You can't quarantine hunger launching\\ virtual spring food drive\end{tabular} &
  18 &
  Food \\ \hline
\begin{tabular}[c]{@{}c@{}}Oil price hits four-year low, touches\\ \$26.20 - @Profitpk\end{tabular} &
  14 &
  Price \\ \hline
\begin{tabular}[c]{@{}c@{}}The workers that society has suddenly decided\\ it can't operate without are often the most\\ poorly-rewarded\end{tabular} &
  19 &
  Worker \\ \hline
\end{tabular}
\end{center}
\end{table}
Picking up a very high peak can sometimes give a little bit of granularity~\citep{Argyrou}. If the same keywords are repeated in more than one topic, it is probably a sign that the 'k' is too big. If the coherence score appears to keep rising, it might be best to choose a model that offers a higher CV before flattery. We have selected a model with 20 topics itself as shown in Figure 2.
\subsection{Selecting Top Unigram Feature form aspect terms cluster:}
Aspect terms of tweets associated with each topic are clustered. Hence, we get 20 aspect terms clusters corresponding to 20 topics generated from the mallet version of LDA. Figure 4 shows the top 20 Unigrams of the topic no. 7 and Figure 5 shows the top 20 Unigrams of the topic no. 6 of COVID-19 tweets. Here, "scam" is the top unigram feature with the highest count for topic no. 7 and "shopping" is the top unigram feature with the highest count for topic no. 6. Hence, "scam" is the topic label corresponding to topic no. 7 and "shopping" is the topic label corresponding to topic no. 6. In this way, the top Unigram feature of each aspect terms cluster is assigned to be the topic label of that particular topic.
\subsection{Qualitative Evaluation of Topics}
We present a set of randomly selected example topics produced by the proposed system for the dataset, as seen in Table 2. Each identified topic is presented with a top Unigram feature from its aspect terms cluster and is matched with a corresponding description of the tweet, as well as a sample tweet is obtained using the keyword of the topic. As shown in Table 2, the labels in the topics found are also closely coherent and well-aligned with tweet descriptions. We can also find out more
useful information about the story of the real world, simply
looking at its topic label.
\begin{figure}
\begin{center}
\includegraphics[width=\textwidth,height=8cm]{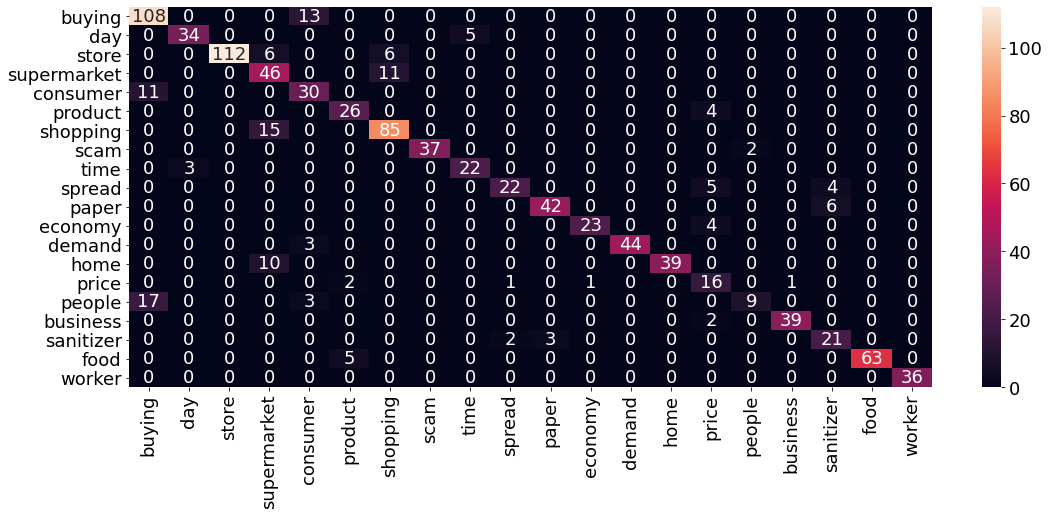}
\caption{Confusion Matrix of proposed model}
\end{center}
\end{figure}
\subsection{Effectiveness Analysis}
In this experiment, we show the effectiveness of our proposed model on the discovered automated topic labels with respect to manually assigned topic labels in terms of accuracy. We select 1000 explainable Tweets randomly from the dataset and assign 20 topic labels generated by the proposed model to those selected Tweets by the annotator. We then compare the two labels in terms of accuracy and our proposed model achieves the accuracy of 85.48\% for the manually assigned topic labels generated by the proposed model. Figure 6 shows the confusion matrix. The observation from the matrix shows that few labels were assigned wrongly. Diverse human perception on labeling the Tweets is the possible reason for incorrect prediction of labels.
\section{Discussion}
Identifying the topic of the tweets of social media platforms such as Twitter can provide meaningful insights into understanding people's ideas, which can be difficult to achieve with traditional strategies, such as manual methods. Overall, we have expanded the analysis to see if we can find the dependency of semantic features of user tweets on various issues on COVID-19-related topics. We have found and detected relevant topics with corresponding topic labels about COVID-19 tweets. Since the LDA is a probabilistic model, when used in the documents, it assumes that each document is made up of a combination of invisible (hidden) topics from a collection, in which the topic is defined as a separate distribution of words. Regarding labeled topics using the Unigram feature of aspect terms in COVID-19 tweets, it is possible to notice several issues related to needs and highlighting conversations of people or users on Twitter.

Overall, the proposed framework enables us to generate significant information from COVID-19 related tweets. We believe that our approach can also be helpful in other domains of applications like healthcare, education, agriculture, cyber-security, etc.  These types of statistical contributions can be useful in determining the positive and negative actions of the public using online platforms like Twitter and gathering user feedback to help researchers and clinicians to understand the behavior of people in critical situations.
\section{Conclusion and Future Work}
In this paper, we have presented an effective framework to find reasonable latent topics with corresponding labels of COVID-19-related issues from Twitter dynamically. Our framework helps to analyze the public opinion on Twitter for COVID-19 after the closure of Wuhan. However,  we believe that the results of this paper will help to understand people's concerns and needs in relation to COVID-19-related issues.

Regarding future work, we plan to test the framework in Bengali language and integrate it with the detection of multiclass sentiment polarities using hybridization of deep learning architectures like CNN, LSTM or BiLSTM~\citep{Sarker3}. We only received information from Twitter, but for people who did not use Twitter to express their views, we could not gather the focus of the topic and their feelings. By increasing COVID-19, we need to increase our data volume to provide answers to broader public opinion and control of relevant departments.
%
%

\end{document}